# A fast and Accurate Sketch Method for Estimating User Similarities over Trajectory Data

Hua Wang

College of Information Science and Engineering, Hunan University, 410082, ChangSha, China

Nest2018@hnu.edu.cn

**Abstract**.

In a complex urban environment, due to the unavoidable interruption of GNSS positioning signals and the accumulation of errors during vehicle driving, the collected vehicle trajectory data is likely to be inaccurate and incomplete. A weighted trajectory reconstruction algorithm based on a bidirectional RNN deep network is proposed. GNSS/OBD trajectory acquisition equipment is used to collect vehicle trajectory information, and multi-source data fusion is used to realize bidirectional weighted trajectory reconstruction. At the same time, the neural arithmetic logic unit (NALU) is introduced into the trajectory reconstruction model to strengthen the extrapolation ability of the deep network and ensure the accuracy of trajectory prediction, which can improve the robustness of the algorithm in trajectory reconstruction when dealing with complex urban road sections. The actual urban road section was selected for testing experiments, and a comparative analysis was carried out with existing methods. Through root mean square error (RMSE, root mean-square error) and using Google Earth to visualize the reconstructed trajectory, the experimental results demonstrate the effectiveness and reliability of the proposed algorithm.

## 1 Introduction

The increasing availability of location and mobility data enables a number of applications, e.g., enhanced navigation services and parking, context-based recommendations, or waiting time predictions at restaurants, which have great potential to improve the quality of life in modern cities. However, the large-scale collection of location data also raises privacy concerns, as mobility patterns may reveal sensitive attributes about users, e.g., home and work places, lifestyles, or even political or religious inclinations. A hands of privacy attacks based on location data, such as de-anonymization[1], social relationship inference attack[2], which rely on similarity estimation. However, they strive to extract features over all the mobility data. Due to the nature of applications involve massive volume of data, it is prohibitive to collect the entire data streams, especially when computational and storage resources are limited.

Recently, due to the proliferation of mobile devices (e.g., smartphones), a new perception paradigm has emerged[3], namely: mobile crowdsensing. With large-scale embedded sensors (such as cameras, GPS and biomedical sensors) and advanced wireless communication technologies (such as WIFI, 4G/5G and Bluetooth) on mobile devices, mobile users/participants can easily use their Mobile devices perform various tasks issued by task owners on crowdsourcing platforms. Due to the inherent advantages of mobile crowd sensing (mobile crowdsensing) applications such as lightweight deployment costs, abundant sensing resources, and large-scale spatio-temporal coverage, it has been applied to environmental monitoring, commerce, intelligent transportation, and auxiliary medical care so far. and many other fields ([4]).

Spatial crowdsourcing (spatial crowdsourcing) [5] as a special group perception (crowdsourcing), so that multiple task owners can outsource their spatio-temporal tasks to the spatial crowdsourcing server, and then the spatial crowdsourcing server will recruit a group of The mobile user completes the task issued by the task owner in the designated sensing location or area. In these spatial crowdsourcing applications, the aggregation and analysis of data submitted by mobile users is more valuable to the task issuer than simply collecting raw sensory data, because many potential values can be mined more directly by aggregating data. , for example: the average rating of a restaurant will show how popular the restaurant is with customers; The maximum driving speed on a road can provide real-time traffic information for road planning [6].

In 2017, Facebook was fined millions of euros by the Spanish Data Protection Commission for alleged violations of EU data protection regulations enforced by Spain. In May 2018, the European Union proposed and implemented the white paper GDRP (General Data Protection Regulation) for sensitive personal information. Although the implementation of GDRP has curbed the unscrupulous violations of user privacy by large companies to a certain extent, it cannot completely eradicate the cancer. At the same time, in July 2018, Facebook was fined $5 billion by the Federal Commission in the United States for using improper means to process users' personal privacy information. It can be seen that the privacy of user data is getting more and more attention worldwide, and it has further reduced the company's trust in the minds of users. In order to prevent similar large-scale privacy leaks in spatial crowdsourcing applications, our analysis found that the following two privacy issues need to be urgently resolved.

There have been various attempts to address the privacy and security issues of spatial crowdsourcing, but they all only focus on task assignment ([9][10]) or data aggregation ([11][12]) privacy protection. In the process of task assignment, in order to protect the location privacy of participants when requesting tasks, fuzzy [11], perturbation ([12][13]) methods are widely used to hide the precise location of users, but this will affect The precision with which tasks are assigned. The method based on encryption [14] to hide or distort the user's precise location can indeed improve the accuracy of the assignment but it cannot protect the privacy of the mobile user's preference for participating in the task.

Although there have been many attempts to protect users' privacy during data aggregation, most existing systems rely on trusted servers to aggregate crowdsourced data. However, the server may be attacked by hackers, or the user's privacy may be deliberately leaked due to interests, so that the server becomes untrusted. Then, an untrusted server should not be allowed to accept or store raw data directly from the user, otherwise the user's personal identity and sensitive information will be exposed. Unfortunately, data aggregation frameworks that rely on trusted servers can no longer protect individual privacy when the server is not trusted. Recently, some work has taken into account the problem of untrusted servers. [15][16] used homomorphic encryption technology to protect user privacy under untrusted servers. Using homomorphic encryption can ensure accurate data aggregation. , but it often results in large computation and communication overhead. [17] used differential privacy technology to ensure that the server can only access perturbed data, but from the research results of [18] it can be seen that aggregated data is extremely vulnerable to membership inference attacks and when sacrificed With the availability of large data, differential privacy can effectively resist member reasoning attacks. In order to overcome these shortcomings, inspired by [30], data privacy and inference privacy are introduced, where data privacy means that the original data uploaded by each user is not observed by the crowdsourcing platform (server), and inference privacy means that the data obtained from the aggregated data is not observed by the crowdsourcing platform (server). Sensitive information about users (for example, whether observations submitted by that user are included in aggregated data) cannot be inferred.

## 2 Architecture of models for estimating user similairties

The challenges including: 1. How to efficiently and accurately allocate tasks while protecting the location privacy of task publishers and mobile users, as well as the preference privacy of mobile users participating in tasks; 2. How to maximize the efficiency of aggregated data while protecting the user's data privacy and reasoning privacy; 3. How to efficiently resist collusion attacks by multiple untrusted third parties.

The overview of our model is comprised of three major steps/components: (1) Correlation preserving representation. A hybrid encoder module that explores three types of encoders: temporal semantic, spatial semantic and contextual semantic to obtain spatial-temporal-semantic representations. (2) Correlation features sampling and learning. (3) Memory-efficient similarity estimation. modeling similarities in pairs of users' trip information.

## 3 Correlation preserving representation

Cloaking includes spatial cloaking and temporal cloaking. Spatial cloaking is based on reducing the precision of users' actual location by sending a generalized region instead of a precise point to the server while the temporal cloaking diminishes the precision in time value. The existence of delay introduced by temporal cloaking makes it cannot work well in LBS which requires high real-time. So, cloaking techniques usually use spatial cloaking or the spatial-temporal cloaking.

Step 1. Use key distribution to initialize the system;

Step 2. Each user registers with GM (group manager) and obtains membership key and task request token;

Step 3. Before assigning the task, the server anonymously verifies the task request token and returns an approval or rejection message to the user;

Step 4. Only users who have received the approval message can request a certain number of pseudonyms from the PA (pseudom authority), through which the perception report can be submitted anonymously;

Step 5. For each perception report received, the server will evaluate its level of trust and issue a receipt with embedded feedback;

Step 6. After all the receipts have been collected, users can fairly redeem rewards from the server. At the same time, a reputation update token is also returned to the user and the GM (group manager) will update the reputation;

Step 7. During the whole process, if the reputation of the user or the trust value of the submitted perception report does not meet the minimum requirements, the corresponding user is deleted from the system, or the perception report submitted by the same user is marked as invalid.

$$q_j = \widetilde{Q_j}(D) - \widetilde{Q_j'}(D)$$

$$= Q_j(D) - Q_j'(D) - static[u \neq x^{(u)}] - dynamic_{Q_j}[u \neq x^{(u)}]$$

$$+ \sum_{i \in A'} dynamic_{Q_j}[a^{(i)} = x^{(i)}] + dynamic_{Q_j}[s = 0]$$

$$- \sum_{i \in A'} dynamic_{Q_j'}[a^{(i)} = x^{(i)}] - dynamic_{Q_j'}[s = 0]$$

$$+ \left(dynamic_{Q_j}[\Delta_j] - dynamic_{Q_j'}[\Delta_j]\right)$$

## 4 Correlation features sampling and learning

Defend against deanonymization attacks and semantic attacks. Semantic attack refers to: the motivation to access a certain address. By generalizing (merging) the positions of two users in time and space dimensions, the generalized position points meet three indicators: k-anonymity (resisting re-identification attacks, ensuring indistinguishability), l-diversity, t -closeness. Among them, the purpose of k-anonymity is to resist re-identification attacks and ensure indistinguishability; l-diversity is to make the POI categories of this location rich enough (not less than l types) to avoid semantic attacks; t-closeness is to make POIs of this location The distribution is not much different from the overall urban POI distribution (KL divergence distance is less than t). Defines the spatio-temporal resolution loss. Definition of optimization problem: Objective - minimize the loss of spatio-temporal resolution, constraints - guarantee k-anonymity, l-diversity, t-closeness.

At present, the privacy protection methods in offline trajectory publishing, such as trajectory clustering, fake trajectory, etc., only regard trajectory data as a sequence of location points with time attributes in Euclidean space, only considering the time and space attributes of trajectories, but ignoring The corresponding position information of each sampling point on the trajectory in the actual environment is obtained, that is, the semantic attribute of the trajectory.

Generally, the location points on the user trajectory can be divided into moving points and staying points. The moving point can only analyze which roads the user has passed, but the staying point can reflect the important location characteristics of the user in a certain period of time. By analyzing the stay points, we can know the places that users frequently visit, and then infer the user's work address, hobbies, even religious beliefs, physical conditions and other private information. Therefore, compared with moving points, staying points will expose more sensitive information of users. Protecting the stay point can not only ensure the privacy of the user, but also reduce the damage to the original trajectory, and achieve a better balance between privacy protection and data availability.

In real life, different users may have different sensitivities to the same semantic location. For example, patients and doctors may have different sensitivities to hospitals. Patients may not want to reveal their physical health, but doctors generally do not mind their work. The location is leaked. Therefore, the user's individual privacy needs cannot be ignored when protecting the trajectory. If the same processing standard is adopted for all users, it may lead to insufficient track protection of some users, resulting in privacy leakage, and excessive track protection of some users, resulting in data loss.

Ignoring the semantic properties of trajectories makes some existing schemes vulnerable to semantic attacks. Compared with the locations they pass by, users are more concerned about whether the places they have visited frequently and stayed for a long time will leak their privacy. Therefore, in order to maintain the maximum integrity of the trajectory, it is not necessary to protect all the sampling points on the trajectory.

$$\tilde{Q}(D) = Q(D) + \sum_{i=1}^{h} static[C_i] + \sum_{i=1}^{h} dynamic_Q[C_i]$$
$$Q \equiv count(a_1 = x_1 \wedge \ldots \wedge a_k = x_k \wedge s = 0),$$
$$Q_1 \equiv count(a_2 = x_2 \wedge \ldots \wedge a_k = x_k \wedge s = 0),$$
$$Q_1' \equiv count(a_1 \neq x_1 \wedge a_2 = x_2 \wedge \ldots \wedge a_k = x_k \wedge s = 0).$$

## 5 Experiment

**5.1 Experimental Setting**

Our social relationship inference framework lies in constructing several measures from discriminative multi-dimensions. Note that we intend to build an elementary but effective attack system to reveal the social relationship privacy leakage in various services or applications that involve the vehicle mobility datasets.

The collected datasets involve trip information for 200 unique vehicles, a total of 504,720 trips during the period from January to September 2019, of which 56,680 trips were recorded in September 2019. The association between drivers can be captured from the dataset. Besides, the prediction mechanism is to input the location data of any pair of users, and then output the similarity of the pair of users to determine whether they have a social connection. This inference does not depend on the similarity score of other user pairs, so the data of 200 users is sufficient for the training and prediction of our model. And the experimental results also prove that the model is effective on this scale of the dataset. To better understanding, we describe the inference model in details as follows:

Let $F_\theta: \mathbb{R}^d \to \mathbb{R}^k$ be a machine learning model with $d$ input features and $k$ output classes, parameterized by weights $\theta$. For an example $z = (\mathbf{x}, y)$ with the input feature $\mathbf{x}$ and the ground truth label $y$, the model outputs a prediction vector over all class labels $F_\theta(\mathbf{x})$ with $\sum_{i=0}^{k-1} F_\theta(\mathbf{x})_i = 1$ and the final prediction will be the label with the largest prediction probability $\hat{y} = \mathrm{argmax}_i F_\theta(\mathbf{x})_i$. For neural networks, the outputs of its penultimate layer are known as logits, and we represent them as a vector $g_\theta(\mathbf{x})$. The softmax function is then computed on logits to obtain the final prediction result $\hat{y}$.

Moreover, there exist related works that use fewer datasets. For example, the authors in [11] collect the location data of 85 volunteers for friendship prediction.

**5.2 Experimental results and analysis**

Figure 1 show the reconstruction trajectory and position error results of various methods in the turning section of the intersection, and the GNSS failure time is 40s. This intersection is similar to a "right-angle turn", and the motion state of the vehicle is similar to that of a right-angle turn. They all enter at a low speed and accelerate to leave after passing the intersection. In this road section, due to frequent speed changes, the acceleration from the motion sensor is inaccurate, and the angular velocity is more accurate, resulting in the shape of the reconstructed trajectory of DR-OBD being similar to the real trajectory but shorter. From Figure 1, it can be seen that the trajectory reconstructed by our proposed Bi-RNN-NALU method is the closest to the real trajectory, especially at the end of the trajectory, which almost coincides with the end of the real trajectory. The effect of the DR-RNN method of only forward trajectory reconstruction is also good. SVR may not be able to correctly fit the features at the end of the trajectory due to excessive cumulative errors, and the direction offset is large, while GPR is almost invalid in the middle and late stages of the trajectory. Trajectory deviation is large. As shown in Figure 1, the position error of the proposed Bi-RNN-NALU method at each trajectory point is always kept at the lowest level, compared with the position error of other methods at the end of the trajectory increases.

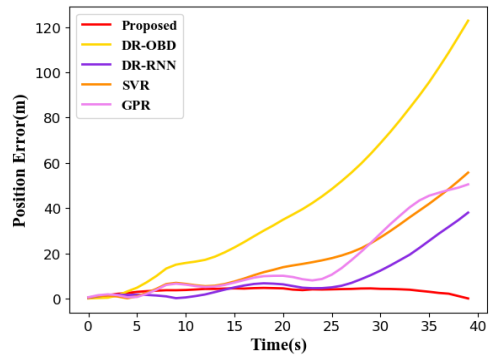

Figure1 Reconstructed Trajectory Position Error

The visualization of the confusion matrix is shown as follow. Accordingly, the true negative rate $TNR = \frac{TN}{TN+FP} = \frac{127}{127+4} \approx 0.969$.

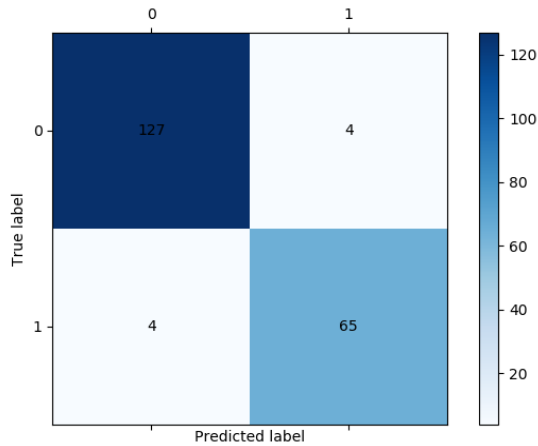

Figure2 The visualization of the confusion matrix

Figure 2 shows the results of reconstructed trajectories and position errors of various methods on straight road sections, and the GNSS failure time is 40 seconds. It can be seen from Figure 1that except for the direction of the SVR reconstructed trajectory changing slightly at the end of the trajectory, the other reconstructed trajectories are basically linear. Combining with Figure 1, it can be seen that DR-RNN performs well, but the two-way. The effect of weighted Bi-RNN-NALU is significantly improved, and the reconstructed trajectory almost coincides with the real trajectory. From the DR-OBD, it can be seen that the inherent noise and error accumulation of the motion sensor in the car on this road section is relatively large, which directly leads to the poor effect of SVR and GPR.

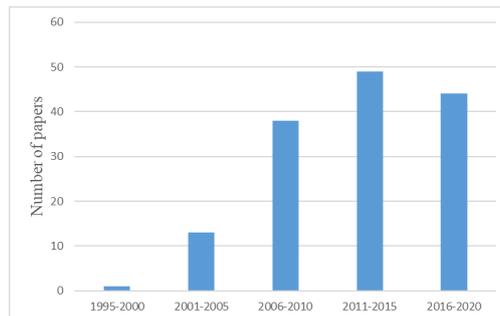

Figure3   Performance at different density.

In order to further demonstrate the performance of the proposed algorithm, the root mean square error (Root Mean Square Error, RMSE) is used as an indicator to verify the effectiveness of each method, and its calculation formula is as follows:

$$RMSE = \sqrt{\frac{\sum_{i=1}^{L}[\hat{s}_i - s_i]^2}{L}}$$

Figure 2 describes the comparison of the root mean square error of the reconstructed trajectory of various methods in the four failure stages. The error results of our proposed algorithm are shown in bold in the table. It can be seen that the Bi-RNN-NALU algorithm can obtain the best accuracy in each road section, because the two-way reconstructed trajectory can use only the forward direction The missing information is reconstructed to ensure the integrity of the trajectory reconstruction. In addition, in the overpass section, the Bi-RNN-NALU and DR-RNN-NALU methods of NALU are greatly improved compared with DR-RNN, which effectively improves the robustness and effectiveness of the algorithm in complex sections such as overpass sections.

## 6 Conclusion

This paper proposes a vehicle trajectory acquisition device based on GNSS/OBD. On this basis, aiming at the problems of inaccuracy and data loss faced by private vehicle trajectory acquisition in urban environment, a new trajectory based on RNN neural network is proposed. Reconstruction algorithm—Bi-RNN-NALU, this algorithm uses RNN as the basic network structure, uses the trajectory information before and after the trajectory is missing, and reconstructs the trajectory from the forward and backward directions respectively, and reconstructs the new vehicle trajectory with weight. Due to making full use of the backward information that the single forward direction does not have, the Bi-RNN-NALU algorithm can reduce the cumulative error of the vehicle trajectory in the later stage and improve the accuracy of the reconstructed trajectory. In addition, in order to meet the challenges of complex road sections, we introduced the NALU joining model to improve the robustness of the algorithm in complex road sections. The accuracy and reliability of the algorithm have been proved by real road tests. Considering the huge amount of trajectory data collected, the next step will be to study the trajectory compression algorithm to improve the efficiency of trajectory data processing.